\documentclass{article}
\usepackage{graphicx}

\usepackage{fvextra} % loads fancyvrb too

\usepackage{fancyvrb}
\usepackage{amsmath} 

\usepackage[numbers,sort&compress]{natbib}

\usepackage{booktabs}
\usepackage{subcaption}

\usepackage{cancel}

\usepackage{authblk}
\usepackage{hyperref}
\title{Extended Joint Models for Longitudinal and Time-to-Event Data: A Tutorial}

\author[1,2]{Pedro Miranda-Afonso\thanks{Corresponding author: \href{mailto:p.mirandaafonso@erasmusmc.nl}{p.mirandaafonso@erasmusmc.nl}}}
\author[1,2]{Dimitris Rizopoulos}

\affil[1]{Department of Biostatistics, Erasmus University Medical Center, The Netherlands}
\affil[2]{Department of Epidemiology, Erasmus University Medical Center, The Netherlands}
\date{}

\newcommand{\codetilde}{\raisebox{-0.6ex}{\texttt{\char`\~}}}
\usepackage{url}
\usepackage{xcolor}
\usepackage{soul}
\definecolor{palegreen}{HTML}{CFFDBC}
\definecolor{paleyellow}{HTML}{FFEE8C}
\definecolor{palered}{HTML}{FF746C}

\usepackage{cancel}

\usepackage{natbib}
\usepackage{bm}

\begin{document}

\maketitle

%\section*{Extended Joint Models for Longitudinal and Time-to-Event Data: A Tutorial}

\begin{abstract}
Shared-parameter joint models for longitudinal and time-to-event data are powerful tools for analyzing repeatedly measured biomarkers, clinical events, and the complex relationships between them. Recent methodological advances have extended the basic framework, which was originally developed for a single event time and a continuous longitudinal biomarker, to more complex scenarios. This tutorial provides a step-by-step guide to fitting extended joint models for longitudinal and time-to-event data using the \textsf{R} package \textbf{JMbayes2}. We cover a range of applications, including models with competing risks, recurrent events, multistate processes, flexible association structures, and multiple longitudinal outcomes following different distributions. Each model is illustrated using simulated data that closely resemble a real-world dataset, with detailed explanations of data organization, model specification, fitting, diagnostics, and interpretation. The tutorial is designed for applied researchers who are interested in analyzing their own data with extended joint models using accessible and reproducible \textsf{R} code.
\end{abstract}

\section{Introduction}

\noindent In medical studies, researchers often monitor patients over time by repeatedly recording biomarker measurements, clinical scores, and events. The events may recur during follow-up, represent transitions through intermediate disease states, or preclude the observation of subsequent events. The longitudinal and event-time processes are often intrinsically related: the longitudinal trajectory of a marker may inform the risk of an event, and the occurrence of an event may induce informative censoring in the longitudinal measurements. As a result, analyzing them separately may lead to loss of information, biased estimates, and inaccurate clinical interpretations.

These problems can be addressed by modeling the joint distribution of the longitudinal and event-time processes, thereby allowing the dependence between them to be explicitly captured. In the widely adopted shared-parameter formulation, the processes are assumed to be conditionally independent given a shared latent structure, which is typically based on subject-specific random effects~\citep{tsiatis2004joint, sousa2011review, rizopoulos2012joint}. Mixed-effects submodels describe the longitudinal outcomes, whereas proportional hazards (relative risk) submodels are often used to describe the corresponding event-time processes. These submodels are linked by specifying the event hazard so that it depends on subject-specific features of the longitudinal process, such as the random effects or the latent subject-specific longitudinal trajectory. This framework has several practical advantages. First, it enables investigators to quantify how features of a subject's longitudinal trajectory relate to the risk of an event. Second, it can be easily extended to more complex settings, such as those with multiple longitudinal outcomes and structured event-time processes. Third, it provides a basis for personalized dynamic prediction, in which individualized conditional risk forecasts are updated as new longitudinal measurements become available during follow-up~\citep{rizopoulos2011dynamic, andrinopoulou2021reflection}.

Early joint models focused on a relatively simple setting: a single Gaussian longitudinal outcome and a single right-censored event time, linked through shared random effects or the latent marker trajectory. More recent research has developed richer and more flexible formulations. Multiple longitudinal outcomes, potentially of different types (e.g., continuous, binary, count), may all be associated with the event of interest~\citep{piulachs2021bayesian, miranda2025joint}. The event process may involve competing risks~\citep{elashoff2008joint, williamson2008joint,andrinopoulou2014joint}, recurrent events~\citep{liu2008analysis, liu2009joint, kim2012joint, krol2016joint}, combinations of both~\citep{miranda2025joint}, or intermediate events~\citep{rizopoulos2024using}. Event times may be left truncated~\citep{piulachs2021bayesian} or interval censored~\citep{yang2025personalized}, while richer association structures can capture, for example, the rate of change of the longitudinal outcome, change over a prespecified time interval, measures of cumulative exposure, or combinations thereof~\citep{mauff2017extension}. For such complex formulations, Bayesian inference provides an efficient framework, particularly when maximum likelihood-based inference requires high-dimensional numerical integration~\citep{rizopoulos2009fully,
rizopoulos2012fast, wang2001jointly, brown2003bayesian,
rizopoulos2011bayesian, alsefri2020bayesian}.

Despite these methodological advances and the increasing availability of software, many applied researchers still face substantial barriers when trying to use extended joint models in practice. These often arise from practical considerations such as how to structure data, how to specify submodels, how to choose and interpret association structures, how to assess and improve the convergence of sampling algorithms, and how to interpret model output in clinically meaningful terms. Previous tutorials have focused on the basic joint model or specific extensions, using different software implementations, rather than providing comprehensive instructions~\citep{asar2015joint, krol2017tutorial,cekic2019tutorial,lovblom2024modeling, baart2021joint}. The \textsf{R} package \textbf{JMbayes2} implements many of these extensions within a common framework, including multiple longitudinal outcomes, recurrent events,
competing risks, multistate processes, and flexible association structures, together with tools for model comparison, diagnostics, and dynamic prediction~\citep{JMbayes2}.

This tutorial provides a practical, step-by-step guide to fitting and interpreting extended joint models using \textbf{JMbayes2}. Our goal is to help applied researchers use these models appropriately and confidently in their work. We begin with a brief overview of the datasets used throughout the tutorial in Section~\ref{sec:datasets} and then present the shared-parameter modeling framework in Section~\ref{sec:modelframe}. Next, in Section~\ref{sec:jmfit}, we fit a basic joint model to establish the workflow in \textsf{R}, covering data organization, syntax, convergence assessment, interpretation, and model comparison. We then extend the framework to multiple longitudinal outcomes, alternative functional forms of association, recurrent events, competing risks, and multistate processes, and combinations of these extensions. We conclude with recommendations for reporting and computational considerations, followed by pointers to further resources.

\section{Datasets} \label{sec:datasets}

\noindent All examples in this tutorial use simulated data designed to mimic the structure of a cystic fibrosis (CF) patient registry used by the authors in previous joint modeling analyses \citep{szczesniak2023lung, afonso2023efficiently, miranda2025joint, miranda2026evaluating, su2026modeling}. This allows us to make the data publicly available alongside the \textbf{JMbayes2} package. Although the simulation parameters were chosen to generate clinically plausible patterns, they are intended solely for tutorial illustration and should not be interpreted as reproducing the natural history, event frequencies, or parameter values of any specific CF cohort. CF is a chronic, multisystem genetic disease characterized by progressive lung damage and malnutrition \citep{farrell2008guidelines}. People with CF often experience recurrent respiratory infections that can lead to irreversible lung damage. Respiratory failure is the principal cause of death, and lung transplantation can substantially improve life expectancy for individuals who develop end-stage lung disease.

The simulated longitudinal outcomes include both Gaussian and non-Gaussian markers, including continuous measures of lung function (e.g., percent predicted forced expiratory volume in 1~s) and binary indicators (e.g., presence of a bacterial infection or CF-related diabetes at a visit). For the event-time component, we consider terminal events (e.g., death), intermediate events (e.g., double-lung transplantation), and recurrent events (e.g., pulmonary exacerbations). Lung transplantation is treated as a non-recurrent event in this tutorial, meaning that each patient can experience at most one transplantation event. 

The complete \textsf{R} code used to generate the simulated datasets and reproduce all examples presented in this tutorial is available in the accompanying GitHub repository: \url{https://github.com/pedromafonso/tutorial_jmbayes2/}.

%The full data simulation procedure is detailed in \hly{Supplementary~Section~C}.

\section{Modeling Framework} \label{sec:modelframe}

Let $i=1,\dots,n$ index subjects, and let $t \geq 0$ denote follow-up time. For a single-event process $k$, let $T_{ik}^*$ be the true event time, let $C_{ik}$ denote an independent right-censoring time, and define the observed event time $T_{ik} = \min(T_{ik}^*, C_{ik})$ and the event indicator $\delta_{ik} = I(T_{ik}^* \leq C_{ik})$. For a recurrent-event process $k$, let $\bm{T}_{ik}^*=(T_{ik1}^*,\ldots,T_{ikL_{ik}}^*)^\top$ denote the vector of true event times, where $T_{ik\ell}^*$ is the true time of the $\ell$th recurrence, $\ell=1,\ldots,L_{ik}$. The corresponding observed recurrence times are $T_{ik\ell}=\min(T_{ik\ell}^*,C_{ik})$, with event indicators $\delta_{ik\ell}=I(T_{ik\ell}^*\le C_{ik})$. 

For subject $i$, let $y_{ij}(t_{ijl})$ be the $j$th ($j=1,\ldots,J$) longitudinal outcome measured at time $t_{ijl}$, $l=1,\ldots,n_{ij}$. The number and timing of longitudinal measurements, as well as the number of recurrent events, may differ between subjects and outcomes. Let $\bm{b}_{ij}$ denote the subject-specific random-effects vector for the $j$th longitudinal outcome, and define
$\bm{b}_i=(\bm{b}_{i1}^\top,\ldots,\bm{b}_{iJ}^\top)^\top$. We assume $\bm{b}_{i} \sim \mathcal{N}(\bm{0}, \bm{D})$, where $\bm{D}$ is the joint covariance matrix of the random effects for all $J$ longitudinal outcomes. These random effects capture latent individual heterogeneity and induce dependence between the longitudinal and event-time processes. The covariance structure $\bm{D}$ creates dependence both within and between the longitudinal processes.

In the shared-parameter framework, the longitudinal and event-time processes are modeled through a set of linked submodels, specified conditionally on the shared latent random effects $\bm{b}_i$, as detailed in Sections~\ref{ssec:m_long}~and~\ref{ssec:m_surv}.

\subsection{Longitudinal Submodel}\label{ssec:m_long}

Conditionally on the random effects $\bm{b}_{ij}$, we assume
\[
y_{ij}(t) \mid \bm{b}_{ij} \sim \mathcal{F}\big\{\mu_{ij}(t), \bm{\phi}_j\big\},
\]
where $\mathcal{F}$ denotes a distribution from an appropriate family, $\mu_{ij}(t)$ is the conditional mean (or another distribution-specific parameter linked to the mean), and $\bm{\phi}_j$ contains additional distributional parameters (e.g., dispersion, shape).

A generalized mixed-effects model specification is obtained through a link function $g(\cdot)$:
\[
g_j\left\{ \mu_{ij}(t) \right\} = \eta_{ij}(t)
= \bm{x}_{ij}^\top(t)\bm{\beta}_j + \bm{z}_{ij}^\top(t)\bm{b}_{ij},
\]
where $\bm{x}_{ij}(t)$ and $\bm{z}_{ij}(t)$ are design vectors for fixed and random effects, respectively, and $\bm{\beta}_j$ is a vector of fixed-effect regression coefficients. The longitudinal linear predictor $\eta_{ij}(t)$ plays a central role in linking the longitudinal and event-time submodels, as detailed in Section~\ref{sec:fforms}.

\subsection{Time-to-Event Submodel}\label{ssec:m_surv}

In line with common practice, the event-time process is modeled through a proportional hazards formulation. Conditionally on the latent random effects and observed covariates, the hazard for subject $i$ at time $t$ is specified as
\[
h_{ik}(t \mid \bm{b}_i, \bm{w}_{ik}(t)) = h_{0k}(t)\exp\left\{
\bm{w}^\top_{ik}(t)\bm{\gamma}_k + \sum_{j}\mathcal{A}_{ijk}(t)
\right\},
\]
where $h_{0k}(t)$ is the baseline hazard function, $\bm{w}_{ik}(t)$ is a vector of baseline or time-varying exogenous covariates with regression coefficients $\bm{\gamma}_k$, and $\mathcal{A}_{ijk}(t)$ is the association term linking the event hazard to the $j$th longitudinal outcome. When multiple functional forms are included for the same longitudinal outcome, we define
\[
\mathcal{A}_{ijk}(t) = \sum_{m=1}^{M_j} \mathcal{A}_{ijkm}(t),
\]
where $\mathcal{A}_{ijkm}(t)$ denotes the $m$th association component for outcome $j$. The baseline hazard may be specified using parametric forms, piecewise-constant functions, or spline-based representations. Left truncation can be accommodated naturally by allowing subjects to enter the risk set only after their entry time $E_{ik} < t$.

To account for additional subject-specific heterogeneity in the event process beyond that explained by the observed endogenous and exogenous covariates and longitudinal process, a frailty term may also be included, for example,
\[
h_{ik}(t \mid \bm{b}_i, \upsilon_i, \bm{w}_{ik}(t)) = h_{0k}(t)\exp\left\{
\bm{w}^\top_{ik}(t)\bm{\gamma}_k + \sum_j\mathcal{A}_{ijk}(t) +  \alpha_{Fk}\upsilon_i
\right\},
\]
where $\upsilon_i\sim \mathcal{N}(0, \sigma_\upsilon^2)$ is a subject-specific frailty term, assumed independent of $\bm{b}_i$. Such a term is particularly useful in recurrent-event settings, where repeated events within the same subject may remain dependent even after conditioning on the longitudinal random effects. For a recurrent-event process, the frailty term $\upsilon_i$ enters the hazard with the coefficient $\alpha_{Fk}$ fixed at 1. In this case, $\upsilon_i$ represents the subject-specific log-frailty shared across recurrences. For other event processes within the same joint model, the same frailty term may also be included with a process-specific coefficient $\alpha_{Fk}$, allowing the strength and direction of the association between the frailty and that event hazard to be estimated from the data.

\subsection{Association Structures} \label{sec:fforms_theory}

\begin{figure}[t]
\includegraphics[width=1.01\textwidth]{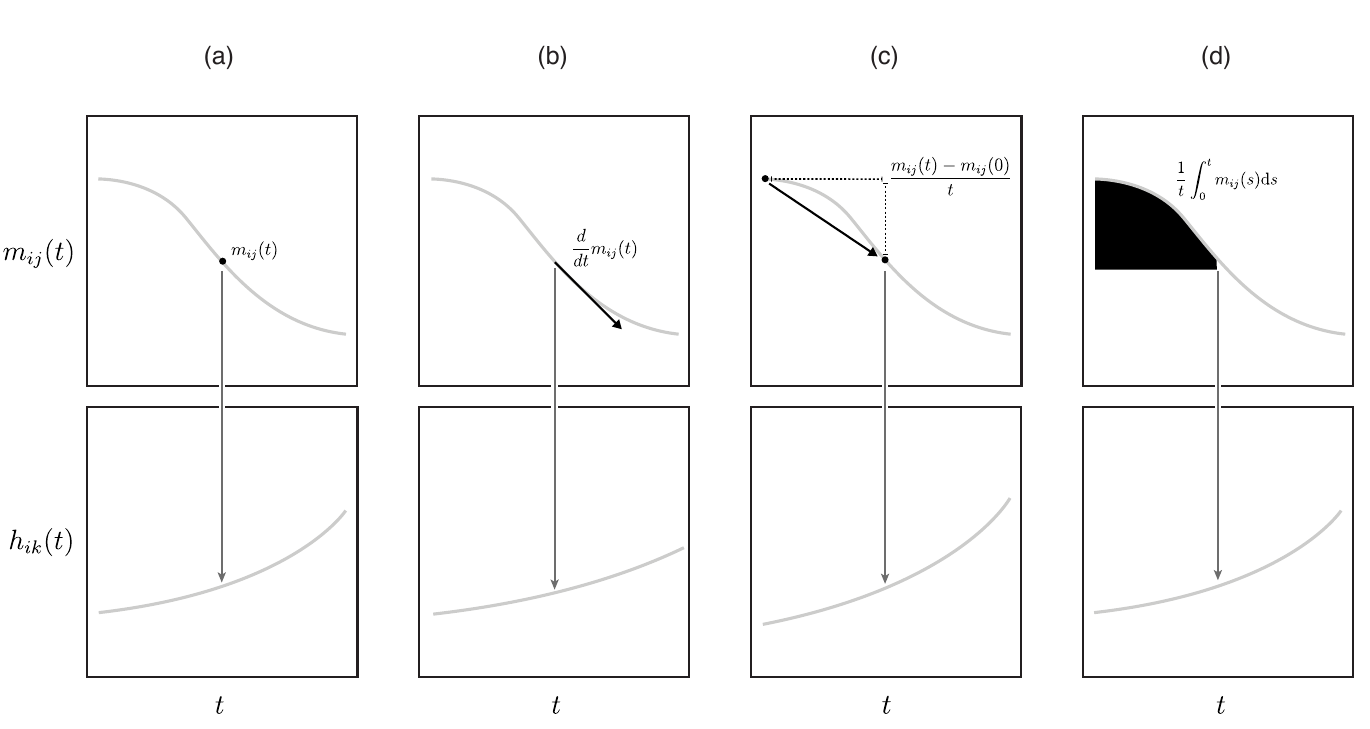}
\caption{Visual representations of association structures that are commonly used to link the latent longitudinal trajectory (top row) to the event hazard (bottom row): (a)~current value, (b)~current slope, (c)~standardized delta, and (d)~standardized area.}
\label{fig:fforms}
\end{figure}

A key component of a shared-parameter joint model is the association structure $\mathcal{A}_{ijk}(t)$, which determines how features of the longitudinal process are related to the event hazard. A common choice is the current-value association (Figure~\ref{fig:fforms}a):
\[
\mathcal{A}_{ijk}(t) = \alpha_{jk} \, m_{ij}(t),
\]
where $m_{ij}(t)$ is either the subject-specific longitudinal linear predictor $\eta_{ij}(t)$ or the expected value $\mu_{ij}(t)$, and $\alpha_{jk}$ is the association parameter. %In this formulation, $\alpha_{jk}$ quantifies the association between the event hazard and the current value of $m_{ij}(t)$, with $\exp(\alpha_{jk})$ yielding the corresponding hazard ratio per unit increase.

In models with an identity link function, the mean and linear predictor coincide; that is, $\mu_{ij}(t)=\eta_{ij}(t)$. With nonidentity link functions, however, the interpretation depends on whether the association is defined on the mean scale or on the linear predictor scale. For example, a current-value association defined on the linear predictor scale, $\mathcal{A}_{ijk}(t)=\alpha_{jk}\,\eta_{ij}(t)$, yields a hazard ratio of $\exp(\alpha_{jk})$ per unit increase in $\eta_{ij}(t)=g\{\mu_{ij}(t)\}$; this depends on the inverse link function and is therefore not constant on the outcome scale. By contrast, if the association is defined on the mean (outcome) scale, $\mathcal{A}_{ijk}(t)=\alpha_{jk}\,\mu_{ij}(t)=\alpha_{jk}\,g^{-1}\{\eta_{ij}(t)\}$, then the hazard ratio for a unit increase in $\mu_{ij}(t)$ is $\exp(\alpha_{jk})$. For this reason, careful attention is needed when specifying and interpreting association structures for binary, count, or bounded longitudinal outcomes.

\begin{figure}[t]
\centering
\includegraphics[width=\textwidth]{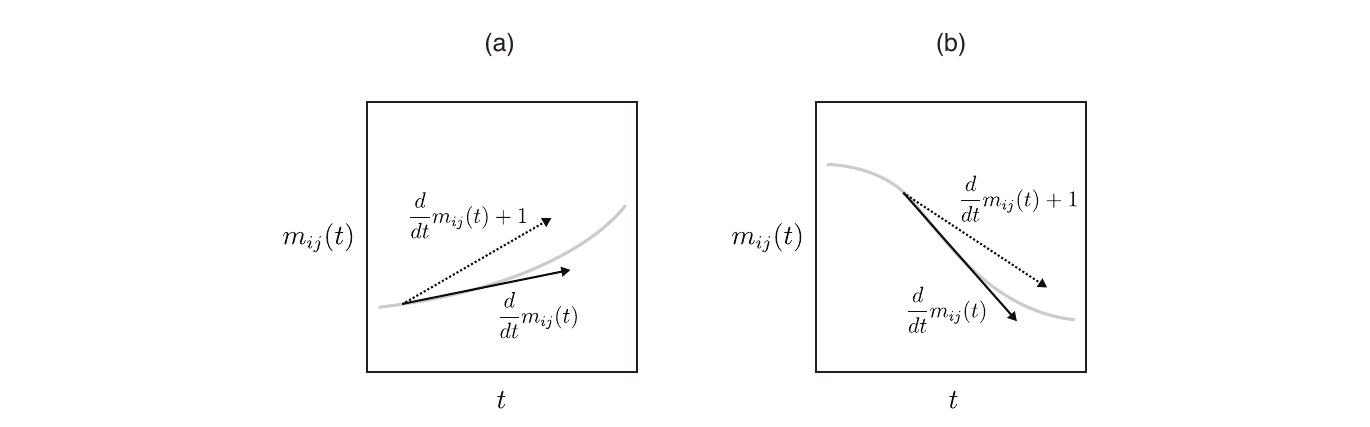}
\caption{Example of current-slope contrasts under two scenarios: (a)~increasing trajectory and (b)~decreasing trajectory, with corresponding slopes that differ by one unit. A unit increase in the trajectory's slope corresponds to a hazard ratio of $\exp(\alpha_{jk})$.}
\label{fig:slope}
\end{figure}

Some applications require richer functional forms. For example, a current-slope association (Figure~\ref{fig:fforms}b) relates the event hazard to the instantaneous rate of change of the underlying longitudinal process. Let $\text{d}m_{ij}(t)/\text{d}t$ denote the first derivative of the longitudinal linear predictor with respect to time. A slope-only association can then be written as
\[
\mathcal{A}_{ijk}(t) = \alpha_{jk} \frac{\text{d}}{\text{d}t}m_{ij}(t).
\]
In this formulation, $\alpha_{jk}$ is the log-hazard ratio associated with a unit increase in the current slope $\text{d}m_{ij}(t)/\text{d}t$. The interpretation depends on the direction of change. When $m_{ij}(t)$ generally increases over time (Figure~\ref{fig:slope}a), a positive $\alpha_{jk}$ implies a higher risk for individuals whose marker is increasing faster, whereas a negative $\alpha_{jk}$ implies a higher risk for individuals whose marker is increasing more slowly. When $m_{ij}(t)$ generally declines over time (Figure~\ref{fig:slope}b), a positive $\alpha_{jk}$ implies a higher risk for individuals whose marker is declining more slowly, whereas a negative $\alpha_{jk}$ implies a higher risk for individuals whose marker is declining faster.

More commonly, the hazard is allowed to depend on both the current value and the current slope:
\[
\mathcal{A}_{ijk}(t) = \alpha_{jk1} m_{ij}(t) + \alpha_{jk2} \frac{\text{d}}{\text{d}t}m_{ij}(t).
\]
Such formulations can be useful when rapid deterioration (or improvement) is clinically relevant in addition to the current marker level. 

It is important to distinguish between the information used to estimate the longitudinal trajectory and the specific longitudinal summary that enters the hazard function. In current-value and current-slope associations, the hazard at time $t$ depends on the value or slope of the latent longitudinal process evaluated at that time point. Although all available longitudinal measurements contribute to the estimation of the subject-specific trajectory, the association parameter itself does not represent an effect of the full marker history up to time $t$. Therefore, association structures based on changes or cumulative summaries over a time window may be preferable in some settings, and they can also be easier to communicate to broader audiences.

One such option is the delta association (Figure~\ref{fig:fforms}c), which relates the hazard function to the change in the longitudinal process over a prespecified time lag $\Delta t$, $\Delta>0$. Fot $t>0$, two common variants are
\[
\mathcal{A}_{ijk}(t) = \alpha_{jk}\{m_{ij}(t)-m_{ij}(t-\min\left\{\Delta t,t\right\})\}
\]
and
\[
\mathcal{A}_{ijk}(t) = \alpha_{jk}\frac{m_{ij}(t)-m_{ij}(t-\min\left\{\Delta t,t\right\})}{\min\left\{\Delta t,t\right\}}.
\]

The first variant captures the absolute change over $(\max\left\{0, t-\Delta t\right\},t]$ (absolute delta), whereas the second corresponds to the average rate of change over the same interval (standardized delta). In both cases, $\alpha_{jk}$ is the log-hazard ratio associated with a unit increase in the corresponding change. The standardized delta is equivalent to the current slope when $m_{ij}(t)$ is linear in time.

Another important association structure is the standardized area (Figure~\ref{fig:fforms}d), in which the hazard function depends on a summary of the subject's longitudinal history up to time $t>0$, as follows:
\[
\mathcal{A}_{ijk}(t) = \alpha_{jk} \frac{1}{t-\max\{0,t-\Delta t\}}\int_{\max\{0, t-\Delta t\}}^t m_{ij}(s)\, \text{d}s,
\]
where $\Delta t$ is a prespecified time lag, $\Delta t > 0$. The standardized integral in this equation yields the average level $m_{ij}(s)$ over the available interval $[\max\{0, t-\Delta t\},\, t]$, facilitating comparisons between subjects with different follow-up durations. In this formulation, $\alpha_{jk}$ is the log-hazard ratio associated with a unit increase in this standardized area. As a special case, when $\Delta t \geq t$, the association depends on the average cumulative exposure from time zero up to time $t$.

\begin{figure}[t]
\centering
\includegraphics[width=\textwidth]{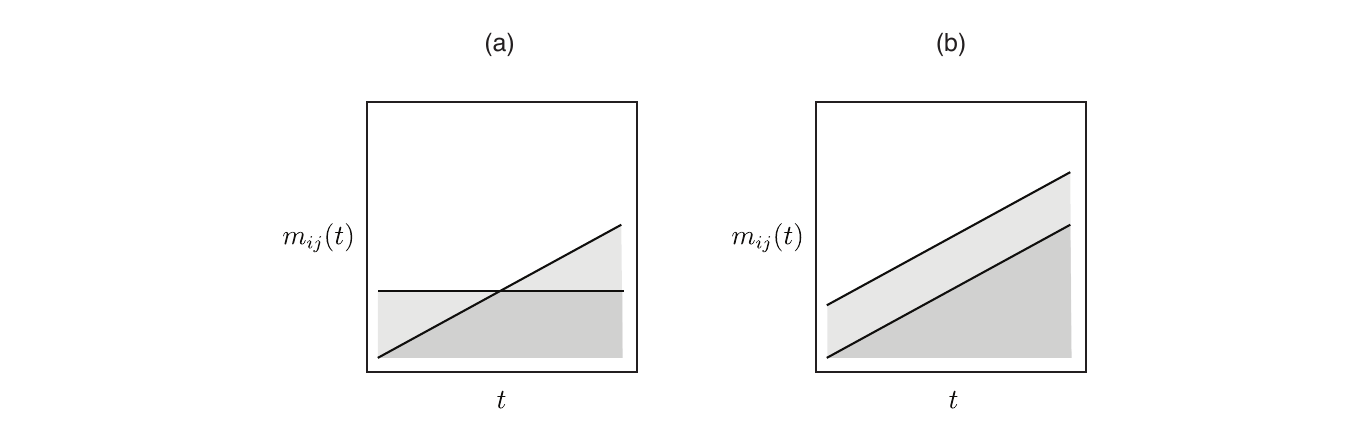}
\caption{Illustration of complementary information captured by standardized delta and standardized area association structures. (a)~Two individual trajectories with the same average exposure over a time window may differ in their change within that window. (b)~Two individual trajectories with the same change over a time window may differ in their average exposure.}
\label{fig:delta_area}
\end{figure}

More generally, combinations of current value, current slope, absolute or standardized delta, and standardized area may be specified to align the model with the relevant research question or hypothesized physiological mechanism. These summaries can provide complementary information about the longitudinal process. For example, a standardized delta can distinguish subjects with the same cumulative exposure over a time window but with different trajectories during that period (Figure~\ref{fig:delta_area}a). Conversely, a standardized area can distinguish subjects with the same change over a time window but different cumulative exposure (Figure~\ref{fig:delta_area}b). 

These functionals can also be interacted with baseline or exogenous covariates $\bm{w}_{ik}(t)$, for example, to allow the association structure to vary across subgroups or over time. The association parameter itself can also be allowed to vary over time, $\alpha_{jk}=\alpha_{jk}(t)$, to capture changes in the strength of association during follow-up as violations of hazard proportionality. In this case, the hazard ratio corresponding to a unit increase in the chosen longitudinal summary becomes time dependent, with $\exp\{\alpha_{jk}(t)\}$ interpreted as the hazard ratio at time $t$~\citep{andrinopoulou2018improved, piulachs2021bayesian}. %A convenient way to present such time-varying associations is through effects plots.%, \hly{as discussed in Section}~\ref{sec:interp}.

In some applications, it may be useful to cap the time at which the longitudinal trajectory enters the association structure so that the event hazard depends only on features of the latent trajectory up to a specified time point, that is, $\mathcal{A}_{ijk}(\min\left\{t,t_i^*\right\})$, where $t_i^*$ denotes the capping time for subject $i$. For example, after an intermediate event such as organ transplantation, the association of pretransplant organ markers with subsequent outcomes may be based on the pretransplant trajectory evaluated only up to the transplantation time \citep{liu2024dynamic}. A similar strategy may be useful when longitudinal measurements are concentrated early in follow-up and extrapolation to distant time points is considered unreliable. In that case, $t_i^*$ can be chosen to restrict the association structure to the part of the longitudinal trajectory supported by the observed data.

\section{Model Fitting} \label{sec:jmfit}

The implementation for any joint model in \textbf{JMbayes2} follows a modular workflow: the longitudinal and event-time submodels are first specified using standard functions from the \textsf{R} packages \textbf{GLMMAdaptive} \citep{GLMMadaptive}, \textbf{nlme} \citep{nlme-package}, and \textbf{survival} \citep{survival-package}, and the resulting fitted objects are then passed to \texttt{jm()} from \textbf{JMbayes2}, which estimates the joint model using Markov chain Monte Carlo (MCMC) algorithms. 

Importantly, although this workflow resembles a two-step procedure, all model parameters are inferred jointly within a single model fit. Therefore, when the longitudinal and event-time processes are associated, the estimates from the joint model are expected to differ from those obtained by fitting the two submodels separately. This modular specification primarily serves to ease joint model specification for users who are already familiar with fitting mixed-effects and Cox models in \textsf{R}. 

Although the event-time submodel is specified using \texttt{coxph()} from \textbf{survival}, the joint model requires an explicit specification of the baseline hazard within \texttt{jm()}. By default, \textbf{JMbayes2} uses penalized B-splines (P-splines) for the baseline hazard, but alternative specifications are also available, including natural cubic splines, piecewise-constant or piecewise-linear functions, and a parametric Weibull hazard.

\subsection{Data Structure}

To fit a joint model in \textbf{JMbayes2}, the user must supply two datasets: one for the longitudinal measurements and one for the event times. The longitudinal dataset has a long format, illustrated in Figure~\ref{fig:datasets}a, with one row for each measurement encounter. Each row contains the subject identifier, the measurement time, the longitudinal outcome(s) value(s), and the baseline or time-varying covariates evaluated at that visit. 

For a basic joint model without time-varying exogenous covariates in the time-to-event submodel, the event-time dataset, as illustrated in Figure~\ref{fig:datasets}b, contains one row for each individual, with the event or right-censoring time, the event indicator (1, observed event; 0, right-censored), the subject identifier, and baseline covariates for the survival submodel. When delayed entry (left truncation), exogenous time-varying covariates, intermediate or recurrent event times are present, the event-time dataset may be organized in start--stop format (Figure~\ref{fig:datasets}c). Each row corresponds to a follow-up interval for a given subject, bounded by the start of that interval and either the time of the next event or the end of follow-up. The dataset may contain multiple rows per subject, with each row corresponding to a follow-up interval over which time-varying covariates are assumed to be constant. 

The same subject identifiers must be used in both longitudinal and event-time datasets, and the longitudinal measurement times must be recorded on the same time scale as the event times (i.e., using the same time unit and a common time zero).

\begin{figure}[t]
\centering
% ---------- Top row: (a) and (b) ----------
\begin{subfigure}[t]{0.56\textwidth}
\centering
\footnotesize
(a)\begin{tabular}{rrrrrr}
\toprule
\texttt{id} & \texttt{time} & \texttt{lf} & \texttt{pa} & \texttt{bc1} & \texttt{tc1} \\
\midrule
\texttt{1} & \texttt{0.20} & \texttt{0.81} & \texttt{NA} & \texttt{0}  & \texttt{8.4} \\
\texttt{1} & \texttt{0.65} & \texttt{0.75} & \texttt{1}  & \texttt{0}  & \texttt{11.2} \\
\texttt{2} & \texttt{0.10} & \texttt{0.90} & \texttt{0}  & \texttt{NA} & \texttt{6.9} \\
\texttt{2} & \texttt{0.21} & \texttt{NA}   & \texttt{0}  & \texttt{1}  & \texttt{7.4} \\
\texttt{2} & \texttt{0.71} & \texttt{0.71} & \texttt{0}  & \texttt{1}  & \texttt{9.1} \\
\multicolumn{6}{c}{$\vdots$} \\
\bottomrule
\end{tabular}
\end{subfigure}\hfill
\begin{subfigure}[t]{0.40\textwidth}
\centering
\footnotesize
(b)\begin{tabular}{rrrrr}
\toprule
\texttt{id} & \texttt{time} & \texttt{event} & \texttt{bc1} & \texttt{bc2} \\
\midrule
\texttt{1} & \texttt{4.30} & \texttt{1} & \texttt{0} & \texttt{2} \\
\texttt{2} & \texttt{3.10} & \texttt{0} & \texttt{1} & \texttt{3} \\
\multicolumn{5}{c}{$\vdots$} \\
\bottomrule
\end{tabular}
\end{subfigure}
\vspace{6pt}

\vspace{6pt}

% ---------- Bottom row: centered (c) ----------
\begin{subfigure}[t]{0.72\textwidth}
\centering
\footnotesize
(c)\begin{tabular}{rrrrrrr}
\toprule
\texttt{id} & \texttt{start} & \texttt{stop} & \texttt{event} & \texttt{bc1} & \texttt{bc2} & \texttt{tc2} \\
\midrule
\texttt{1} & \texttt{0.25} & \texttt{1.20} & \texttt{0} & \texttt{0} & \texttt{2} & \texttt{0.62} \\
\texttt{1} & \texttt{1.20} & \texttt{4.30} & \texttt{1} & \texttt{0} & \texttt{2} & \texttt{0.70} \\
\texttt{2} & \texttt{0.10} & \texttt{3.10} & \texttt{0} & \texttt{1} & \texttt{3} & \texttt{0.81} \\
\multicolumn{7}{c}{$\vdots$} \\
\bottomrule
\end{tabular}
\end{subfigure}
\caption{Examples of data structures used in \textbf{JMbayes2}. (a) Longitudinal dataset. (b) Event-time dataset with one row per subject. (c) Event-time dataset in start--stop format. Abbreviations: \texttt{bc1}, baseline covariate 1 (e.g., sex); \texttt{bc2}, baseline covariate 2 (e.g., genotype class); \texttt{lf}, longitudinal marker 1; \texttt{id}, patient identifier; \texttt{pa}, longitudinal marker 2; \texttt{time}, time since index time; \texttt{tc1}, time-varying covariate 1 (e.g., air pollution exposure); \texttt{tc2}, exogenous time-varying covariate 2 (e.g., air humidity level).} \label{fig:datasets}
\end{figure}

This separation into two datasets reflects how the submodels are specified in \textbf{JMbayes2}. The survival submodels are fitted to the event-time dataset, and each mixed-effects submodel is fitted to the same longitudinal dataset. Careful preparation of these two datasets is essential. Although inconsistencies in subject identifiers are typically flagged by error messages, misspecification of the time scale or miscoding of the event indicator may not be detected and can lead to incorrect results without warnings or error messages.

\subsection{Basic Syntax} \label{sec:basic_jm}

To lay the groundwork for the extended joint models that are the main focus of this tutorial, we first fit a basic joint model using \textbf{JMbayes2}. In this first example, we consider a single continuous longitudinal outcome (e.g., lung function), denoted by \texttt{lf}, and a single event-time outcome that may be right-censored (e.g., time to the composite endpoint of transplantation or death), as follows:
\begin{equation}
\label{eq:basic_jm}
\begin{cases}
\texttt{lf}_i(t) &\sim \mathcal{N} \{m_i(t), \sigma_y^2\} \\
m_i(t)
&= \beta_{0}
 + \beta_{1}t
 + \beta_{2}\texttt{sex}_i
 + \beta_{3}\texttt{ageD}_i
 + \beta_{4}t\,\texttt{sex}_i
 + b_{0i}
 + b_{1i}t,\\
h_i(t)
&= h_0(t)\exp\left\{
\gamma_1\texttt{sex}_i
+ \gamma_2\texttt{ageD}_i
+ \alpha m_i(t)
\right\},
\end{cases}
\end{equation}
where $(b_{0i},b_{1i})^\top\sim \mathcal{N}(\boldsymbol 0,\boldsymbol D)$. The first step is to specify the two submodels separately. For the longitudinal outcome, we specify a linear mixed-effects model using \texttt{lme()} from \textbf{nlme}, and for the event-time outcome, we specify a proportional hazards model using \texttt{coxph()} from \textbf{survival}. In the longitudinal submodel, the fixed-effects part includes follow-up time \texttt{time} (years), a binary covariate \texttt{sex} (female vs.\ male) and its interaction with time, and a continuous covariate \texttt{ageD} denoting age at diagnosis, whereas the random-effects part includes a subject-specific intercept and slope for follow-up time. In the time-to-event submodel, \texttt{stop} denotes the time (years) of death or transplantation, whichever occurs first, or, for right-censored individuals, the last time they were known to be alive and without transplantation, and \texttt{status} is the event indicator. The two submodels may include different covariates. Introductions to modeling longitudinal and event-time outcomes can be found in \cite{nlme-book} and \cite{survival-book}, respectively.
\fvset{fontsize=\small}
\begin{Verbatim}[breaklines, breakanywhere, commandchars=\\\{\}]
> install.packages("JMbayes2")
> library("JMbayes2")

> fit_long1 <- lme(lf \codetilde{} time * sex + ageD, random =\codetilde{} time | id,
                   data = long)

> fit_surv <- coxph(Surv(stop, status) \codetilde{} sex + ageD, data = surv)
\end{Verbatim}

\noindent Finally, the fitted submodels are combined using \texttt{jm()}. The argument \texttt{time\_var} identifies the follow-up time variable in the longitudinal dataset \texttt{long}. 

\fvset{fontsize=\small}
\begin{Verbatim}[breaklines, breakanywhere, commandchars=\\\{\}]
> fit_jm <- jm(Surv_object = fit_surv, Mixed_objects = fit_long1,
               time_var = "time")
\end{Verbatim}

\noindent When calling the function \texttt{jm()}, the user can adjust MCMC settings, such as the number of iterations (\texttt{n\_iter}) and burn-in (\texttt{n\_burnin}), and pass additional control options via the \texttt{control} argument. Further details are available in the function documentation (see \texttt{?jm}).

Bayesian inference requires prior distributions for all unknown model parameters and the random effects. In \textbf{JMbayes2}, default hyperparameters are chosen to be weakly informative and to yield stable estimation in typical applications, while remaining sufficiently flexible for a wide range of data settings. The default prior specification is detailed in Supplementary~Section~A. Users can modify any of these choices via the \texttt{priors} argument to \texttt{jm()}. The priors used in a fitted model can be inspected directly from the fitted object using the syntax \texttt{fit\_jm\$priors}.%, where \texttt{fit\_jm} is a fitted joint model object.

\subsubsection{Model Convergence Checks} \label{sec:converg}

Because \textbf{JMbayes2} uses MCMC estimation, its output must be assessed with convergence diagnostics as part of the model fitting process. A first step is to inspect trace plots of the MCMC samples (i.e., sampled parameter values across iterations). Stable, well-mixed chains without visible trends are generally taken as evidence of convergence. %Figure~\ref{fig:traceplot} shows an example trace plot for the association coefficient (\texttt{alphas}). 
The call \texttt{traceplot(fit\_jm, parm = "alphas")} (or \texttt{ggtraceplot()}) generates the trace plot for the association coefficient (\texttt{alphas}). A list of all model parameter names is provided in Supplementary~Table~\ref{tab:supp:model_params}.

%\begin{Verbatim}[breaklines, breakanywhere, commandchars=\\\{\}]
%> traceplot(fit_jm, parm = "alphas") \textcolor{gray}{\# or ggtraceplot(fit_jm, parm = "alphas")}
%\end{Verbatim}

%\begin{figure}[t]
%\centering
%\includegraphics[width=\textwidth]{Figures/placeholder.png}
%\caption{Example trace plot for the association parameter in a joint model. It shows the sampled values across iterations for each MCMC chain. Convergence is indicated by stable traces with good mixing and substantial overlap across chains, without apparent trends or shifts in location.}
%\label{fig:traceplot}
%\end{figure}

\noindent Some other useful diagnostics are posterior density plots and the Gelman--Rubin convergence diagnostic $\hat{R}$ \citep{gelman1992inference}, which can be obtained using \texttt{densplot()} (or \texttt{ggdensityplot()}) and \texttt{gelman\_diag()}, respectively. The $\hat{R}$ value is also reported by the \texttt{summary()} method, discussed in the next section. Values of $\hat{R}$ close to 1 (many authors suggest an upper threshold of 1.1 \citep{gelman2013bayesian}) indicate good mixing across chains. In practice, convergence assessment should consist of graphical inspection, comparison across chains, and a check that posterior summaries are stable under longer runs. When these diagnostics suggest that convergence has not been attained, the number of MCMC iterations and burn-in iterations should be increased. If convergence remains unsatisfactory after longer runs, this may indicate that the specified model is too complex for the information available in the data, in which case a more parsimonious model specification should be considered. These checks should precede model comparison and interpretation.

\subsubsection{Interpretation of Model Output} \label{sec:interp}

Once the model has been fitted and satisfactory MCMC convergence has been achieved, the next step is to examine the parameter estimates. We illustrate the information returned by \texttt{summary()} for the fitted joint model object \texttt{fit\_jm}. The output is organized into several blocks.

\fvset{fontsize=\small}
\begin{Verbatim}
> summary(fit_jm)
\end{Verbatim}
\begin{Verbatim}[frame=single, rulecolor=\color{gray}]
Call:
JMbayes2::jm(Surv_object = fit_surv, Mixed_objects = fit_long1, 
             time_var = "time")
\end{Verbatim}
%\vspace{-17pt}

\begin{Verbatim}[frame=single, rulecolor=\color{gray}]
Data Descriptives:
Number of groups: 500		Number of events: 330 (66%)
Number of observations:
  lf: 5833
\end{Verbatim}
%\vspace{-17pt}

\begin{Verbatim}[frame=single, rulecolor=\color{gray}]
                  DIC      WAIC     LPML
marginal    -10796.55 -10864.10 5163.303
conditional -16252.17 -16549.87 7901.688
\end{Verbatim}
%\vspace{-17pt}

\begin{Verbatim}[frame=single, rulecolor=\color{gray}]
Random-effects covariance matrix:
                    
       StdDev   Corr 
(Intr) 0.1736 (Intr) 
time   0.0137 -0.0608
\end{Verbatim}
%\vspace{-17pt}

\begin{Verbatim}[frame=single, rulecolor=\color{gray}]
Survival Outcome:
             Mean  StDev    2.5%   97.5%      P   Rhat
sex        0.1792 0.1380 -0.0870  0.4438 0.2000 1.0110
ageD       0.0450 0.0341 -0.0226  0.1097 0.2018 1.0002
value(lf) -1.0952 0.3141 -1.6994 -0.4779 0.0004 1.0030
\end{Verbatim}
%\vspace{-17pt}

\begin{Verbatim}[frame=single, rulecolor=\color{gray}]
Longitudinal outcome: lf (family = gaussian, link = identity)
               Mean  StDev    2.5%   97.5%      P   Rhat
(Intercept)  0.8469 0.0162  0.8149  0.8786 0.0000 1.0006
time        -0.0302 0.0011 -0.0324 -0.0279 0.0000 1.0045
sex         -0.0430 0.0161 -0.0749 -0.0115 0.0071 1.0008
ageD        -0.0113 0.0049 -0.0208 -0.0018 0.0207 0.9999
time:sex    -0.0028 0.0016 -0.0058  0.0003 0.0793 1.0016
sigma        0.0604 0.0006  0.0592  0.0616 0.0000 1.0044
\end{Verbatim}
%\vspace{-17pt}

\begin{Verbatim}[frame=single, rulecolor=\color{gray}]
MCMC summary:
chains: 3 
iterations per chain: 3500 
burn-in per chain: 500 
thinning: 1 
time: 4 sec
\end{Verbatim}

\noindent The \texttt{Call} block reports the function call and the objects supplied to \texttt{jm()}. The \texttt{Data Descriptives} block summarizes the sample size (\texttt{Groups}), the number of events observed during follow-up, and the number of longitudinal measurements (\texttt{Observations}) used in fitting. The third block reports information criteria that are useful for model comparison: the deviance information criterion (DIC), the Watanabe--Akaike information criterion (WAIC), and the log pseudo-marginal likelihood (LPML). These criteria differ between marginal and conditional formulations because they treat the subject-specific random effects differently, with the marginal formulation integrating them out and the conditional formulation conditioning on them \citep{ariyo2020bayesian}. Next, the \texttt{Random-effects covariance matrix} block provides posterior summaries for the random-effects standard deviations and their correlation.

The \texttt{Survival Outcome} block reports posterior summaries for the event-time regression coefficients, including the association term (here, \texttt{value(lf)}). All event-time regression coefficients are reported on the log-hazard ratio scale. The columns \texttt{Mean} and \texttt{StDev} provide the posterior mean and standard deviation, and \texttt{2.5\%} and \texttt{97.5\%} define a 95\% credible interval. The column \texttt{P} reports a posterior tail-area probability for the corresponding parameter (often used as an analogue of a two-sided $p$-value), and the \texttt{Rhat} column contains the $\hat{R}$ value. For the current-value association structure used here, $\exp(-1.0952)$ can be interpreted as the hazard ratio associated with a unit increase in the longitudinal linear predictor. 

The \texttt{Longitudinal Outcome} block reports posterior summaries for the longitudinal fixed effects (on the linear predictor scale) and additional distributional parameters (e.g., dispersion or shape). Here, \texttt{sigma} is the standard deviation of the conditional distribution of $\texttt{lf}_i(t)$.

The \texttt{MCMC summary} block records the number of chains, iterations, burn-in, thinning, and elapsed time.

%\noindent Beyond numerical summaries, it is often useful to visualize the non-linear effects of covariates. The \hly{\texttt{effects()}} method can be used to produce marginal plots of selected covariate relationships, for example \texttt{effects(fit\_jm)}. \hly{Supplementary~Figure~SX} shows some of the marginal plots obtained for this model.

\subsubsection{Model Comparison}~\label{sec:mcomp}

\noindent When several plausible joint models are under consideration, model comparison can help assess whether added complexity is supported by the data. \textbf{JMbayes2} provides the function \texttt{compare\_jm()}, which compares fitted models using the information criteria described in Section~\ref{sec:interp}.

Consider the following alternative  specification for $m_i(t)$ in the longitudinal submodel of Equation~\ref{eq:basic_jm}:
\[
m_i(t)= \beta_0
 + \beta_1 t
 + \beta_2\texttt{sex}_i
 + \beta_3 B_1(\texttt{ageD}_i)
 + \beta_4 B_2(\texttt{ageD}_i)
 + \beta_5 t\,\texttt{sex}_i
 + b_{0i}
 + b_{1i}t,
\]
where $B_1(\cdot)$ and $B_2(\cdot)$ denote two natural cubic spline basis functions for age at diagnosis, replacing the previous linear effect. The event-time submodel retains the same form. The \texttt{update()} function is used to refit \texttt{fit\_long1} by changing only its fixed-effects formula; the random-effects structure and dataset are retained.
\begin{Verbatim}[breaklines, breakanywhere, commandchars=\\\{\}]
> fit_long1_b <- update(fit_long1, fixed = lf \codetilde{} time * sex + ns(ageD, 2))

> fit_jm_b <- jm(Surv_object = fit_surv, Mixed_objects = fit_long1_b,
                 time_var = "time")

> compare_jm(fit_jm, fit_jm_b, type = "marginal")

               DIC      WAIC     LPML
  fit_jm -10796.55 -10864.10 5163.303
fit_jm_b -10631.55 -10242.24 4705.378

The criteria are calculated on the basis of the marginal log-likelihood.
\end{Verbatim}

\noindent In this example, the lower DIC and WAIC and higher LPML for \texttt{fit\_jm} favor the simpler model with a linear effect of age at diagnosis. Model comparison should include subject-matter considerations, as models with alternative formulations imply different hypotheses about the longitudinal and event-time processes and the mechanisms that link them.

\subsection{Joint Model with Multiple Longitudinal Outcomes} \label{sec:jm_mlong}

%Many applications involve more than one longitudinal outcome. %In CF registry settings, for example, lung function and nutritional status are both recorded during follow-up visits, which often occur at irregular and outcome-specific times. 
Joint models with multiple longitudinal outcomes allow the event-time process to depend on multiple biomarkers simultaneously, while accounting for correlation between them through correlated random effects. Different types of outcomes may be modeled using different distributions and link functions. At the time of writing, \textbf{JMbayes2} supports mixed-effects submodels with Gaussian, Student’s-t, gamma, beta, unit-Lindley, censored normal, binomial, Poisson, negative binomial, and beta-binomial outcomes.

For illustration, we extend the joint model fitted in the previous example by adding a binary marker, coded in the dataset as \texttt{pa}, which could represent, for example, whether a respiratory culture is found to be positive for \textit{Pseudomonas aeruginosa} at a given visit. The resulting joint model can be written as follows:
\[
\begin{cases}
\texttt{lf}_i(t) &\sim \mathcal{N} \{m_{i1}(t), \sigma_y^2\} \\
m_{i1}(t)
&= \beta_{10}
 + \beta_{11}t
 + \beta_{12}\texttt{sex}_i
 + \beta_{13}\texttt{ageD}_i
 + \beta_{14}t\,\texttt{sex}_i
 + b_{10i}
 + b_{11i}t,\\
\texttt{pa}_i(t)
&\sim \operatorname{Bernoulli}\{\pi_i(t)\},\\
\operatorname{logit}\{\pi_i(t)\}
&=m_{i2}(t)
= \beta_{20}
 + \beta_{21}t
 + \beta_{22}\texttt{sex}_i
 + \beta_{23}\texttt{ageD}_i
 + b_{20i}
 + b_{21i}t,\\
h_i(t)
&= h_0(t)\exp\left\{
\gamma_1\texttt{sex}_i
+ \gamma_2\texttt{ageD}_i
+ \alpha_1 m_{i1}(t)
+ \alpha_2 m_{i2}(t)
\right\}.
\end{cases}
\]
where $(b_{10i},b_{11i},b_{20i},b_{21i})^\top \sim \mathcal{N}(\boldsymbol 0,\boldsymbol D)$. We model \texttt{pa} as a Bernoulli outcome, using a mixed-effects logistic regression with a logit link function. The fixed-effects part includes follow-up time \texttt{time} and the covariates \texttt{sex} and \texttt{ageD}, whereas the random-effects part includes a subject-specific intercept and slope for follow-up time. For non-Gaussian outcomes, we use the function \texttt{mixed\_model()} from \textbf{GLMMAdaptive} and then pass the fitted mixed-effects objects to \texttt{jm()} as a list through the argument \texttt{Mixed\_objects}.

The longitudinal information for multiple outcomes must be provided in a single long-format dataset (Figure~\ref{fig:datasets}a). Different outcomes may be observed at different times, so some outcomes may be missing (\texttt{NA}) at a given measurement occasion. The event-time dataset remains as in Section~\ref{sec:basic_jm}.

\fvset{fontsize=\small}
\begin{Verbatim}[breaklines, breakanywhere, commandchars=\\\{\}]
> fit_long2 <- mixed_model(pa \codetilde{} time + sex + ageD, random =\codetilde{} time | id,
                           family = binomial(link = "logit"),
                           data = long)

> fit_jm2 <- jm(Surv_object = fit_surv,
                Mixed_objects = list(fit_long1, fit_long2),
                time_var = "time")
\end{Verbatim}

\fvset{fontsize=\small}
\newpage
\begin{Verbatim}[breaklines, breakanywhere, commandchars=\\\{\}]
> summary(fit_jm2)
...
Survival outcome:
             Mean  StDev    2.5%   97.5%      P   Rhat
sex        0.1588 0.1358 -0.1029  0.4239 0.2484 1.0016
ageD       0.0308 0.0341 -0.0357  0.0974 0.3707 1.0024
value(lf) -1.0757 0.3260 -1.7280 -0.4404 0.0004 1.0062
value(pa)  0.3160 0.1170  0.0884  0.5480 0.0020 1.0142
...
\end{Verbatim}
With multiple longitudinal outcomes, the event-time submodel includes at least one association parameter for each outcome. These association parameters should be interpreted conditionally on the other longitudinal summaries and covariates in the submodel. 

When an outcome is modeled with a nonidentity link function, as in the \texttt{pa} example shown here, the corresponding default association is defined on the linear predictor scale, and its interpretation should reflect this scale. %For example, \texttt{value(pa)} corresponds to a current-value association defined through the subject-specific linear predictor of the binary outcome. 
Thus, the reported mean value of \texttt{value(pa)} ($0.3160$) is the logarithm of the hazard ratio associated with a unit increase in the subject-specific log-odds, not the probability, of a positive culture being detected at time $t$.

Convergence issues may arise when the number of random effects is large due to the number of outcomes or nonlinear time effects. These issues may be related to high posterior correlation among variance components or weak information for certain random-effects terms. In such cases, convergence may be improved by increasing the number of iterations, considering more informative priors, and adopting more parsimonious random-effects structures or covariance specifications when appropriate. The argument \texttt{which\_independent} to \texttt{jm()} can be used to enforce prior independence between selected longitudinal outcomes.

\subsection{Functional Forms of Association} \label{sec:fforms}

%Section~\ref{sec:modelframe} introduced the definitions of the main association structures that link the longitudinal and event-time processes in the shared-parameter framework.
\begin{sloppypar}
In the previous two examples, we used the \texttt{jm()} default current-value association for the longitudinal outcomes on the linear predictor scale. The \texttt{functional\_forms} argument to \texttt{jm()} can be used to specify alternative association structures for specific outcomes. Multiple functionals can also be specified for the same outcome. As discussed in Section~\ref{sec:fforms_theory}, the choice of association structure should be guided by the scientific question. Table~\ref{tab:mth:forms} summarizes the functional forms and associated transformation functions available in \textbf{JMbayes2}.
\end{sloppypar}

\begin{table}[h!]
\centering
\caption{Functional forms available in \textbf{JMbayes2} to link the longitudinal and time-to-event outcomes, with the associated transformation functions.}
\label{tab:mth:forms}
\resizebox{\textwidth}{!}{%
\begin{tabular}{lll} 
\toprule
\textbf{Functional form} & \textbf{Function} & \textbf{Argument} \\
\midrule

Current value
& $m_{ij}(t)$
& \texttt{value($\cdot$)} \\

& $\log\left\{m_{ij}(t)\right\}$
& \texttt{vlog(value($\cdot$))} \\

& $\log_2\left\{m_{ij}(t)\right\}$
& \texttt{vlog2(value($\cdot$))} \\

& $\log_{10}\left\{m_{ij}(t)\right\}$
& \texttt{vlog10(value($\cdot$))} \\

& $\sqrt{m_{ij}(t)}$
& \texttt{vsqrt(value($\cdot$))} \\

& $\exp\left\{m_{ij}(t)\right\}$
& \texttt{vexp(value($\cdot$))} \\

& $\exp\{m_{ij}(t)\}/[1+\exp\{m_{ij}(t)\}]$
& \texttt{vexpit(value($\cdot$))} \\

& $m_{ij}^2(t)$
& \texttt{poly2(value($\cdot$))} \\

& $m_{ij}^3(t)$
& \texttt{poly3(value($\cdot$))} \\

& $m_{ij}^4(t)$
& \texttt{poly4(value($\cdot$))} \\

& & \\

Current slope
& $\text{d}\,m_{ij}(t)/\text{d}t$
& \texttt{slope($\cdot$)}\textsuperscript{\textdagger} \\

& $\left|\text{d}\,m_{ij}(t)/\text{d}t\right|$
& \texttt{vabs(slope($\cdot$))} \\

& $\text{d}\exp\left\{m_{ij}(t)\right\}/\text{d}t$
& \texttt{Dexp(slope($\cdot$))} \\

& $\text{d}\operatorname{expit}\left\{m_{ij}(t)\right\}/\text{d}t$
& \texttt{Dexpit(slope($\cdot$))} \\

& & \\

Acceleration
& $\text{d}^2m_{ij}(t)/\text{d}t^2$
& \texttt{acceleration($\cdot$)} \\

& & \\

Standardized area
& $(t-\max\{0,t-\Delta t\})^{-1}
\int_{\max\{0,t-\Delta t\}}^t m_{ij}(s)\,\text{d}s$
& \texttt{area($\cdot$)} \\

Standardized delta
& $\{m_{ij}(t)-m_{ij}(t-\min\{\Delta t,t\})\} /
\min\{\Delta t,t\}$
& \texttt{Delta($\cdot$)} \\

Absolute delta
& $m_{ij}(t)-m_{ij}(t-\min\{\Delta t,t\})$
& \texttt{Delta($\cdot$, standardise = F)} \\

\bottomrule
\end{tabular}
}
\par
\raggedright\footnotesize{
\textsuperscript{\textdagger} \texttt{velocity($\cdot$)} can be used as an alias for
\texttt{slope($\cdot$)}.\par}
\end{table}

\begin{sloppypar}
The joint model \texttt{fit\_jm2} in Section~\ref{sec:jm_mlong} used the default current-value association for both longitudinal outcomes, which is equivalent to specifying \texttt{functional\_forms =\codetilde{} value(lf) + value(pa)} in the \texttt{jm()} call. In this example, we illustrate how to include additional functionals and how to transform an association when the longitudinal outcome uses a nonidentity link function. Retaining the same longitudinal submodels, we modify the event-time submodel to depend on the current value and slope of \texttt{lf} and on the current probability of \texttt{pa}, as follows:
\end{sloppypar} 
\[
h_i(t)
= h_0(t)\exp\left\{
\gamma_1\texttt{sex}_i
+ \gamma_2\texttt{ageD}_i
+ \alpha_1 m_{i1}(t)
+ \alpha_2 \frac{\mathrm{d}}{\mathrm{d}t}m_{i1}(t)
+ \alpha_3 \pi_i(t)
\right\}.
\]
This association structure is implemented by updating the
\texttt{functional\_forms} argument of the previously fitted joint model.
\fvset{fontsize=\small}
\begin{Verbatim}[breaklines, breakanywhere, commandchars=\\\{\}]
> fit_jm3 <- update(fit_jm2, functional_forms =\codetilde{} value(lf) + slope(lf) + vexpit(value(pa)))                    
\end{Verbatim}

\noindent In this specification, \texttt{value(lf)} and \texttt{slope(lf)} enter the hazard simultaneously, yielding separate association coefficients for the current value and current slope, respectively. For the binary outcome \texttt{pa} modeled with a logit link, the term \texttt{vexpit(value(pa))} transforms the subject-specific linear predictor to the outcome scale, so the corresponding association coefficient relates the hazard to the current predicted probability, rather than the log-odds.% that $y_2(t)=1$. %Specifically, \texttt{vexpit(value(pa))} evaluates $\operatorname{expit}\{\eta_{ij}(t)\}=\mu_{ij}(t)$.

In the output produced by \texttt{summary(fit\_jm3)}, the current-slope association term \texttt{slope(lf)} can be interpreted as the log-hazard ratio associated with a unit increase in the instantaneous rate of change of \texttt{lf}, conditional on the other terms in the event-time submodel. The coefficient \texttt{vexpit(value(pa))/10}, $1.5362/10$, is interpreted as the log-hazard ratio associated with a 0.1 increase in the predicted probability, conditional on the other terms in the time-to-event submodel.
\newpage
\fvset{fontsize=\small}
\begin{Verbatim}[breaklines, breakanywhere, commandchars=\\\{\}]
> summary(fit_jm3)
...
Survival outcome:
                     Mean  StDev     2.5%   97.5%      P   Rhat
sex                0.1481 0.1382  -0.1170  0.4196 0.2904 1.0027
ageD               0.0340 0.0346  -0.0351  0.1010 0.3187 1.0002
value(lf)         -0.9073 0.3364  -1.5756 -0.2527 0.0058 1.0119
slope(lf)         -9.1843 6.4236 -21.7486  3.3469 0.1513 1.0278
vexpit(value(pa))  1.5362 0.5157   0.5202  2.5412 0.0051 1.0264
...
\end{Verbatim}

%When interpreting association parameters, it is important to keep track of the scale on which the functional form is defined. For the binary outcome $y_2$ modeled with a logit link, specifying \texttt{value(pa)} links the hazard to a current value defined on the log-odds (linear predictor) scale, whereas \texttt{vexpit(value(pa))} corresponds to the same current value mapped to the probability scale. In practice, transforming to the outcome scale is often helpful for communicating results.

Functional forms can also be interacted with baseline or time-varying exogenous covariates to allow the association between the longitudinal process and the event hazard to vary across groups. For example, specifying \texttt{functional\_forms =\codetilde{} value(lf) + value(lf):sex} estimates sex-specific current-value association parameters for the longitudinal outcome \texttt{lf}.

Alternatively, we can use the standardized area and delta associations described in Section~\ref{sec:fforms_theory}. For example, using a time window of 0.5, the event-time submodel can be written as 
\[
h_i(t)
= h_0(t)\exp\left\{
\gamma_1\texttt{sex}_i
+ \gamma_2\texttt{ageD}_i
+ \alpha_1 A_i(t)
+ \alpha_2 D_i(t)
+ \alpha_3 \pi_i(t)
\right\}.
\]
where
\[
\begin{aligned}
A_i(t)
&=
\frac{1}{t-\max\{0,t-0.5\}}
\int_{\max\{0,t-0.5\}}^t m_{i1}(s)\,\mathrm{d}s,\\
D_i(t)
&=
\frac{
m_{i1}(t)-m_{i1}\{t-\min(0.5,t)\}
}{
\min(0.5,t)
}.
\end{aligned}
\]
\begin{sloppypar}
\noindent This specification is implemented using \texttt{area()} and \texttt{Delta()} in the \texttt{functional\_forms} argument. Both functions allow a window length to be set using the argument \texttt{time\_window}, expressed on the same time scale as \texttt{time\_var}. For \texttt{Delta()}, the argument \texttt{standardise} controls whether the change is divided by the window length, in which case \texttt{Delta()} is interpreted as an average rate of change instead of an absolute change. 
\end{sloppypar}
\fvset{fontsize=\small}
\begin{Verbatim}[breaklines, breakanywhere, commandchars=\\\{\}]
> fit_jm4 <- update(fit_jm2,
                    functional_forms =\codetilde{} area(lf, time_window = 0.5) +
                      Delta(lf, time_window = 0.5, standardise = TRUE) +
                      vexpit(value(pa)),
                    n_iter = 7000L, n_burnin = 1000L, n_thin = 2L)
\end{Verbatim}
%\newpage
%
For this and the more complex models that follow, we increased the number of MCMC iterations and burn-in iterations based on the convergence diagnostics described in Section~\ref{sec:converg}.

\newpage

\fvset{fontsize=\small}
\begin{Verbatim}[breaklines, breakanywhere, commandchars=\\\{\}]
> summary(fit_jm4)
...
Survival Outcome:
                    Mean  StDev     2.5%   97.5%      P   Rhat
sex               0.1450 0.1385  -0.1208  0.4111 0.3147 1.0013
ageD              0.0349 0.0357  -0.0356  0.1044 0.3224 1.0000
area(lf, ...)    -0.9088 0.3418  -1.6014 -0.2447 0.0098 1.0115
Delta(lf, ...)   -9.6961 6.4881 -22.5184  2.6530 0.1304 1.0100
vexpit(value(pa)) 1.5347 0.5726   0.4563  2.6983 0.0044 1.0308
...
\end{Verbatim}

\noindent \begin{sloppypar}The terms \texttt{area(lf, time\_window = 0.5)} and \texttt{Delta(lf, time\_window = 0.5, standardise = TRUE)} summarize the average level and rate of change of the underlying trajectory of the outcome over the preceding half time unit, respectively. As in the previous examples, their coefficients are interpreted as log-hazard ratios per unit increase in the corresponding summary quantity, conditional on the other terms in the submodel.\end{sloppypar}

An \texttt{IE\_time} argument can also be passed to the functions defining the functional forms to cap the time at which the longitudinal trajectory enters the association structure. This argument sets a subject-specific time point, such as the time of an intermediate event (IE), up to which the longitudinal functional is evaluated, as discussed in Section~\ref{sec:fforms_theory}.

\subsection{Joint Model with Recurrent Events} \label{sec:jm_rec}

In many applications, it is of interest to study how one or more longitudinal markers are related to an event that may recur multiple times for the same individual during follow-up. In CF, for example, pulmonary exacerbations may occur repeatedly over time, and their risk may be associated with the underlying lung function trajectory. Unlike the event considered in the preceding examples, recurrent events do not terminate follow-up: an individual may experience several events and remain at risk of further occurrences until they die or are censored. \textbf{JMbayes2} supports joint models with a recurrent event process and allows the recurrent-event risk intervals to be defined on either the calendar or gap time scale. The calendar time scale uses study entry as the time origin, whereas the gap time scale resets the time origin after each recurrent event. The gap time scale is often natural when studying how risk evolves after the most recent event, whereas the calendar time scale is preferable when risk is primarily driven by the time since study entry. A detailed discussion of these two time scales and their implications for recurrent-event joint models is provided elsewhere~\citep{miranda2025joint}.

For illustration, we retain the longitudinal submodel from Section~\ref{sec:basic_jm} and use a gap-time specification for the recurrent-event process. The corresponding recurrent-event hazard can be written as
\[
h_i(t)
= h_0(t-T_{i,\ell-1})\exp\left\{
\gamma_1\texttt{sex}_i
+ \gamma_2\texttt{ageD}_i
+ \alpha m_i(t)
+ \upsilon_i
\right\},
\]
where $\upsilon_i\sim\mathcal{N}(0,\sigma_\upsilon^2)$ and $t-T_{i,\ell-1}$ denotes the time since the previous recurrent event while subject $i$ is at risk for the $\ell$th recurrence, with $T_{i0}=0$.

Recurrent-event joint models require the event-time dataset to be organized in start--stop format (Figure~\ref{fig:datasets}c). %Each row corresponds to a follow-up interval for a given subject, bounded by the start of that interval (\texttt{start}) and either the time of the next event or the end of follow-up (\texttt{stop}). 
A binary event indicator, \texttt{status}, records whether a recurrent event occurred at the end of the at-risk interval bounded by \texttt{start} and \texttt{stop}. This structure also accommodates nonrisk periods, that is, intervals during which the subject is not at risk of experiencing a new recurrent event, which are represented by gaps between consecutive at-risk intervals for the same individual. It is important to verify that the start--stop intervals correctly represent the periods during which the subject is at risk (excluding any nonrisk periods) and that all event-time variables use the same time scale and time zero as the longitudinal measurements. The longitudinal dataset remains unchanged from the preceding subsections.

\fvset{fontsize=\small}
\begin{Verbatim}[breaklines, breakanywhere, commandchars=\\\{\}]
> surv_rc[1:5, c("id", "start", "stop", "status")]
  id    start      stop status
1  1 0.000000  3.752981      0
2  2 0.000000  1.208693      1
3  2 1.208693  4.445554      1
4  2 4.445554  5.108401      1
5  2 5.108401 10.000000      0
\end{Verbatim}

\noindent In this example, subject 1 experienced no recurrent events and was censored at time 3.75. Subject 2 experienced three recurrent events at times 1.21, 4.45, and 5.11 and was subsequently censored at time 10.

After constructing the event-time dataset, the event process can be specified using a \texttt{coxph()} model with \texttt{Surv(start, stop, status)}. The joint model is then fitted by passing this object to \texttt{jm()}, with the \texttt{recurrent} argument specifying the time scale for the recurrent-event baseline hazard (\texttt{"gap"} or \texttt{"calendar"}). Regardless of the selected time scale, the \texttt{start} and \texttt{stop} variables supplied in the dataset must be defined relative to the same time zero as the longitudinal measurements; the gap-time reindexing is handled internally by \textbf{JMbayes2}.
\fvset{fontsize=\small}
\begin{Verbatim}[breaklines, breakanywhere, commandchars=\\\{\}]

> fit_surv_rc <- coxph(Surv(start, stop, status) \codetilde{} sex + ageD, 
                       data = surv_rc)

> fit_jm_rc <- jm(Surv_object = fit_surv_rc, Mixed_objects = fit_long1, 
                  time_var = "time", recurrent = "gap")
\end{Verbatim}

\fvset{fontsize=\small}
\begin{Verbatim}[breaklines, breakanywhere, commandchars=\\\{\}]
> summary(fit_jm_rc)
...
Frailty standard deviation:
                Mean   2.5%  97.5%
sigma_frailty 0.5609 0.4008 0.7317

Survival Outcome:
             Mean  StDev    2.5%   97.5%      P   Rhat
sex        0.2521 0.1228  0.0078  0.4820 0.0393 1.0029
ageD       0.0177 0.0321 -0.0466  0.0814 0.5851 1.0031
value(lf) -0.6172 0.2578 -1.1238 -0.1216 0.0098 1.0060
...
\end{Verbatim}

\noindent The event process now includes a frailty term that captures within-subject dependence across events and accounts for heterogeneity in event rates beyond that explained by the longitudinal trajectory and observed covariates. The new \texttt{Frailty standard deviation} block reports the estimated standard deviation of the frailty distribution $\sigma_\upsilon$. A value close to zero would indicate little residual heterogeneity beyond what is explained by the longitudinal outcomes and covariates.

\subsection{Joint Model with Competing Risks}  \label{sec:jm_cr}

Competing risks arise when a subject may experience one of several mutually exclusive event types (i.e., when the occurrence of one type of event prevents the occurrence of the others). Joint modeling of the cause-specific event processes allows each event type to be associated with the longitudinal trajectory and accounts for competing events when estimating event probabilities. %Treating one of the events as simple censoring implies that it occurs independently of the other event and longitudinal processes, an assumption that will not hold in many applications. A competing-risks joint model instead allows the cause-specific hazards for the different event types to be modeled simultaneously. 

In our CF example, we consider lung transplantation and death before transplantation as the two event types. Strictly speaking, this setting constitutes a semi-competing risks problem because transplantation is a nonterminal event: death precludes a subsequent transplantation, whereas a transplanted individual may still die afterward. For the purpose of illustrating the competing-risks formulation, however, we treat transplantation as terminating follow-up, so that transplantation and death before transplantation are mutually exclusive events in this analysis.

Retaining the longitudinal submodels from Section~\ref{sec:jm_mlong}, for cause
$k\in\{\mathrm{tx},\mathrm{dth}\}$, the cause-specific hazard is
\[
h_{ik}(t)
= h_{0k}(t)\exp\left\{
\gamma_{1k}\texttt{sex}_i
+ \gamma_{2k}\texttt{ageD}_i
+ \alpha_{1k}m_{i1}(t)
+ \alpha_{2k}\pi_i(t)
\right\}.
\]
In a competing-risks joint model, the event-time dataset must be restructured to the cause-specific stacked format, in which the event-time dataset is expanded to include one row per subject per competing risk. Starting from an event-time dataset \texttt{surv\_cr0} with one row per subject and a multicategory \texttt{status} indicator with levels corresponding to the possible outcomes---\texttt{"alv"} for alive, \texttt{"tx"} for transplantation, and \texttt{"dth"} for death---this conversion can be performed with \texttt{crisk\_setup()}.
\fvset{fontsize=\small}
\begin{Verbatim}[breaklines, breakanywhere, commandchars=\\\{\}]
> surv_cr0[1:3, c("id", "stop", "status")]
  id       stop status
1  1  3.7529815    dth
2  2 10.0000000    alv
3  3  0.1730942    dth
\end{Verbatim}
\newpage
\fvset{fontsize=\small}
\begin{Verbatim}[breaklines, breakanywhere, commandchars=\\\{\}]
> surv_cr <- crisk_setup(surv_cr0, statusVar = "status",
                         censLevel = "alv", nameStrata = "proc")

> surv_cr[1:6, c("id", "stop", "status2", "proc")]
    id       stop status2 proc
1    1  3.7529815       1  dth
1.1  1  3.7529815       0   tx
2    2 10.0000000       0  dth
2.1  2 10.0000000       0   tx
3    3  0.1730942       1  dth
3.1  3  0.1730942       0   tx
\end{Verbatim}

\noindent The resulting \texttt{surv\_cr} stacked dataset contains two rows per subject (one for each cause), with a cause indicator \texttt{proc} and a binary event indicator for that cause \texttt{status2}. The longitudinal dataset is unchanged from Section~4.2.

\begin{sloppypar}For the event process, we fit a cause-specific \texttt{coxph()} model. Using \texttt{strata(proc)} yields a separate baseline hazard for each cause. Covariate effects can differ between causes (using the \texttt{*} or \texttt{:} operator) or be shared (by omitting the interaction). In addition, to allow each longitudinal functional to have a cause-specific association parameter, we interact its functional form with \texttt{proc} in the \texttt{functional\_forms} argument in \texttt{jm()}.\end{sloppypar}
\fvset{fontsize=\small}
\begin{Verbatim}[breaklines, breakanywhere, commandchars=\\\{\}]
> fit_surv_cr <- coxph(Surv(stop, status2) \codetilde{} (sex + ageD):strata(proc),
                       data = surv_cr)

> fit_jm_cr <- jm(Surv_object = fit_surv_cr,
                  Mixed_objects = list(fit_long1, fit_long2),
                  time_var = "time",
                  functional_forms =\codetilde{} (value(lf) + vexpit(value(pa))):proc,
                  n_iter = 7000L, n_burnin = 1000L, n_thin = 2L)
\end{Verbatim}

\fvset{fontsize=\small}
\begin{Verbatim}[breaklines, breakanywhere, commandchars=\\\{\}]
> summary(fit_jm_cr)
...
Survival Outcome:
                             Mean  StDev    2.5%   97.5%      P   Rhat
sex:strata(proc)tx         0.1396 0.2038 -0.2567  0.5334 0.4969 1.0032
sex:strata(proc)dth        0.1748 0.1687 -0.1478  0.5211 0.2956 1.0028
ageD:strata(proc)tx        0.0645 0.0471 -0.0295  0.1535 0.1733 1.0029
ageD:strata(proc)dth      -0.0036 0.0494 -0.1011  0.0919 0.9522 1.0053
value(lf):proctx          -1.3795 0.4380 -2.2289 -0.5313 0.0042 1.0084
value(lf):procdth         -0.7346 0.4462 -1.5935  0.1167 0.1016 1.0169
vexpit(value(pa)):proctx   1.0559 0.7841 -0.4582  2.5867 0.1858 1.0191
vexpit(value(pa)):procdth  1.8087 0.7698  0.3321  3.3409 0.0207 1.0092
...
\end{Verbatim}

\begin{sloppypar}
\noindent The \texttt{Survival Outcome} block from the \texttt{summary(fit\_jm\_cr)} output reports cause-specific regression coefficients. Rows labeled \texttt{proctx} (or \texttt{strata(proc)tx}) and \texttt{procdth} (or \texttt{strata(proc)dth}) correspond to the two competing cause-specific terminal hazards. For example, \texttt{sex:strata(proc)tx} is the sex coefficient, $\gamma_{1\mathrm{tx}}$, for the cause-specific hazard of transplantation, whereas \texttt{value(lf):procdth} represents the current-value association between the biomarker and the cause-specific hazard for death, $\alpha_{1\mathrm{dth}}$. All coefficients are reported on the log-hazard ratio scale, and so applying $\exp(\cdot)$ yields the corresponding cause-specific hazard ratio.
\end{sloppypar}

Cause-specific hazard ratios from a competing-risks model should not be interpreted as marginal risks of each event in the presence of the other. For example, a lower hazard for transplantation does not necessarily imply a lower cumulative incidence of transplantation, because changes in the cause-specific hazards for one event affect exposure to the other.

\subsection{Joint Model with Multistate Processes} \label{sec:jm_ms}

\noindent In some applications, subjects may experience intermediate events before experiencing a terminal event. These intermediate events may alter the risk of subsequent events. Each possible transition is modeled using a transition-specific proportional hazards submodel. For example, a subject with CF may start alive without lung transplantation, undergo transplantation, and subsequently die. The possible transitions are transplantation before death (\texttt{alv-tx}), death without prior transplantation (\texttt{alv-dth}), and death after transplantation (\texttt{tx-dth}). %The association between a longitudinal process and the event process may differ across these transitions.

For this multistate example, we retain the longitudinal submodel specifications from Section~\ref{sec:jm_mlong}, while allowing longitudinal follow-up to continue after transplantation. For transition $k\in\{\mathrm{alv\mbox{-}tx},\mathrm{alv\mbox{-}dth},
\mathrm{tx\mbox{-}dth}\}$, the transition-specific hazard is
\[
h_{ik}(t)
= h_{0k}(t)\exp\left\{
\gamma_{1k}\texttt{sex}_i
+ \gamma_{2k}\texttt{ageD}_i
+ \alpha_{1k}m_{i1}(t)
+ \alpha_{2k}\pi_i(t)
\right\}.
\]

In \textbf{JMbayes2}, multistate processes are specified using an event-time dataset in a start--stop, transition-specific stacked format. Each row corresponds to an interval during which a subject is at risk for a particular transition and contains the \texttt{start} and \texttt{stop} times of the interval, an event indicator \texttt{status}, and a transition indicator \texttt{proc} with levels \texttt{alv-tx}, \texttt{alv-dth}, and \texttt{tx-dth}. Unlike in the competing-risks example, the number of rows may differ between individuals because only transitions that are possible from a subject's current state are included. For example, an individual who has not undergone transplantation will not have a row corresponding to the \texttt{tx-dth} transition.

\fvset{fontsize=\small}
\begin{Verbatim}[breaklines, breakanywhere, commandchars=\\\{\}]
> surv_ms[1:4, c("id", "start", "stop", "status", "proc")]
  id start      stop status    proc
1  1     0  3.752981  FALSE  alv-tx
2  1     0  3.752981   TRUE alv-dth
3  2     0 10.000000  FALSE  alv-tx
4  2     0 10.000000  FALSE alv-dth
\end{Verbatim}

The longitudinal dataset differs from that used in the preceding examples. There, longitudinal follow-up ends at death, transplantation, or censoring. For the multistate model, measurements collected after transplantation are retained because subjects remain under observation and may subsequently experience the \texttt{tx-dth} transition. Thus, \texttt{long} contains no longitudinal measurements after transplantation, whereas \texttt{long\_ms} retains subsequent measurements until death or censoring.

\fvset{fontsize=\small}
\begin{Verbatim}[breaklines, breakanywhere, commandchars=\\\{\}]
> long[long$id == 6, c("id", "time", "lf", "pa")]
   id time        lf pa
91  6 0.00 0.8079272  0
92  6 0.59 0.8537566  0

> long_ms[long_ms$id == 6, c("id", "time", "lf", "pa")]
   id time        lf pa
91  6 0.00 0.8079272  0
92  6 0.59 0.8537566  0
93  6 1.18 0.8248261  0
94  6 1.76 0.8576534  0
95  6 2.35 0.8360533  0
...
\end{Verbatim}

The event-time submodel is specified using \texttt{coxph()} with transition-specific baseline hazards. As in the competing-risks example, transition-specific parameters are obtained by interacting the longitudinal functional or covariates with the \texttt{proc} variable. The longitudinal submodels are refitted using \texttt{long\_ms} to incorporate the additional post-transplant measurements.

\fvset{fontsize=\small}
\begin{Verbatim}[breaklines, breakanywhere, commandchars=\\\{\}]
> fit_long1_ms <- update(fit_long1, data = long_ms)

> fit_long2_ms <- update(fit_long2, data = long_ms)

> fit_surv_ms <- coxph(Surv(start, stop, status) \codetilde{} (sex + ageD):strata(proc),
                       data = surv_ms)

> fit_jm_ms <- jm(Surv_object = fit_surv_ms,
                  Mixed_objects = list(fit_long1_ms, fit_long2_ms),
                  time_var = "time",
                  functional_forms =\codetilde{} (value(lf) + vexpit(value(pa))):proc,
                  n_iter = 14000L, n_burnin = 2000L, n_thin = 4L)
\end{Verbatim}
%\newpage
\fvset{fontsize=\small}
\begin{Verbatim}[breaklines, breakanywhere, commandchars=\\\{\}]
> summary(fit_jm_ms)
...
Survival outcome:
                                 Mean  StDev    2.5%   97.5%   P   Rhat
sex:strata(proc)alv-tx         0.1363 0.2249 -0.3085  0.5758 ... 1.0020
sex:strata(proc)alv-dth        0.1841 0.1677 -0.1493  0.5059 ... 1.0057
sex:strata(proc)tx-dth         0.1417 0.2163 -0.2766  0.5702 ... 1.0021
ageD:strata(proc)alv-tx        0.0641 0.0456 -0.0266  0.1491 ... 1.0031
ageD:strata(proc)alv-dth       0.0016 0.0481 -0.0956  0.0918 ... 1.0024
ageD:strata(proc)tx-dth       -0.0064 0.0640 -0.1346  0.1194 ... 1.0063
value(lf):procalv-tx          -1.4400 0.4302 -2.3483 -0.6539 ... 1.0085
value(lf):procalv-dth         -0.6893 0.4145 -1.4673  0.1228 ... 1.0059
value(lf):proctx-dth          -1.3117 0.5886 -2.4865 -0.1868 ... 1.0106
vexpit(value(pa)):procalv-tx   0.8889 0.6663 -0.4190  2.1564 ... 1.0184
vexpit(value(pa)):procalv-dth  1.9301 0.7877  0.4186  3.5336 ... 1.0012
vexpit(value(pa)):proctx-dth   2.2969 0.7720  0.8703  3.8619 ... 1.0147
...
\end{Verbatim}

\noindent For example, \texttt{value(lf):procalv-dth} is the log-hazard ratio associated with a unit increase in the current value of \texttt{lf} for the transition \texttt{alv-dth}, that is, death before transplantation. \textbf{JMbayes2} can accommodate multiple longitudinal outcomes and more than three states.

\subsection{Joint Model with Competing Risks and Recurrent Events}

Sections~\ref{sec:jm_cr} and~\ref{sec:jm_rec} treated competing risks and recurrent events as separate extensions. However, in many applications, recurrent events are additionally subject to one or more terminating events, which end the follow-up and censor the recurrent-event process. Combining both extensions within a single model avoids the assumption that the competing risks are independent of the recurrent process or that terminal events are noninformative about future recurrences. %Within the framework of Section~\ref{sec:modelframe}, this subsection considers an extended model in which the single time-to-event submodel is replaced with multiple submodels, each sharing the same frailty $\upsilon_i$.

Retaining the longitudinal submodels introduced in
Section~\ref{sec:jm_mlong}, the recurrent-event component is extended to accommodate both longitudinal outcomes:
\[
h_{iR}(t)
= h_{0R}(t-T_{i,\ell-1})\exp\left\{
\gamma_{1R}\texttt{sex}_i
+ \gamma_{2R}\texttt{ageD}_i
+ \alpha_{1R}m_{i1}(t)
+ \alpha_{2R}\pi_i(t)
+ \upsilon_i
\right\},
\]
where $\upsilon_i\sim\mathcal{N}(0,\sigma_\upsilon^2)$ and $t-T_{i,\ell-1}$ denotes the time since the previous recurrent event while subject $i$ is at risk for the $\ell$th recurrence, with $T_{i0}=0$. For terminal process
$k\in\{\mathrm{Tdth},\mathrm{Ttx}\}$, the corresponding cause-specific hazard is
\[
h_{ik}(t)
= h_{0k}(t)\exp\left\{
\gamma_{1k}\texttt{sex}_i
+ \gamma_{2k}\texttt{ageD}_i
+ \alpha_{1k}m_{i1}(t)
+ \alpha_{2k}\pi_i(t)
+ \alpha_{Fk}\upsilon_i
\right\}.
\]
To fit this model in \textbf{JMbayes2}, the event-time information is organized in a single start--stop stacked dataset. For each individual, the resulting dataset contains one row per recurrent at-risk interval and one row per terminal event process. A stratum variable \texttt{proc} distinguishes the different processes.

The event indicator \texttt{status} equals 1 when the corresponding event occurs at the end of the interval and 0 otherwise. If the recurrent and terminal events are recorded in separate datasets, both can be combined using the helper function \texttt{rc\_setup()}, as shown below. In this example, the levels \textsf{R}, \texttt{Tdth}, and \texttt{Ttx} denote the recurrent pulmonary-exacerbation process, the terminal death process, and the terminal transplantation process, respectively. The longitudinal dataset remains unchanged.

\newpage

\fvset{fontsize=\small}
\begin{Verbatim}[breaklines, breakanywhere, commandchars=\\\{\}]
> surv_cr0[1:3, c("id", "stop", "status")]
  id       stop status
1  1  3.7529815    dth
2  2 10.0000000    alv
3  3  0.1730942    dth

> surv_rc[1:6, c("id", "start", "stop", "status")]
  id    start       stop status
1  1 0.000000  3.7529815      0
2  2 0.000000  1.2086929      1
3  2 1.208693  4.4455539      1
4  2 4.445554  5.1084012      1
5  2 5.108401 10.0000000      0
6  3 0.000000  0.1730942      0

> surv_comb <- rc_setup(rc_data = surv_rc, trm_data = surv_cr0,
                        idVar = "id", statusVar = "status",
                        startVar = "start", stopVar = "stop",
                        trm_censLevel = "alv",
                        nameStrata = "proc", nameStatus = "status")

> surv_comb[1:12, c("id", "start", "stop", "status", "proc")]
   id    start       stop status proc
1   1 0.000000  3.7529815      0    R
2   1 0.000000  3.7529815      1 Tdth
3   1 0.000000  3.7529815      0  Ttx
4   2 0.000000  1.2086929      1    R
5   2 1.208693  4.4455539      1    R
6   2 4.445554  5.1084012      1    R
7   2 5.108401 10.0000000      0    R
8   2 0.000000 10.0000000      0 Tdth
9   2 0.000000 10.0000000      0  Ttx
10  3 0.000000  0.1730942      0    R
11  3 0.000000  0.1730942      1 Tdth
12  3 0.000000  0.1730942      0  Ttx
\end{Verbatim}
After constructing the event-time dataset, the event process can be specified via a stratified \texttt{coxph()} model that allows a separate baseline hazard for each stratum. Covariate and biomarker effects can be specified separately for each stratum through interactions with the stratum variable \texttt{proc}. %Similarly, to allow process-specific association parameters, the relevant functional forms can be interacted with the corresponding stratum.
\fvset{fontsize=\small}
\begin{Verbatim}[breaklines, breakanywhere, commandchars=\\\{\}]
> fit_surv_comb <- coxph(Surv(start, stop, status) \codetilde{} (sex + ageD):strata(proc),
                         data = surv_comb)

> fit_jm_comb <- jm(Surv_object = fit_surv_comb, 
                    Mixed_objects = list(fit_long1, fit_long2), 
                    time_var = "time", recurrent = "gap",
                    functional_forms =\codetilde{} (value(lf) + vexpit(value(pa))):proc,
                    n_iter = 14000L, n_burnin = 2000L, n_thin = 4L)
\end{Verbatim}
\fvset{fontsize=\small}
\begin{Verbatim}[breaklines, breakanywhere, commandchars=\\\{\}]
> summary(fit_jm_comb)
...
Survival outcome:
                              Mean  StDev    2.5%   97.5%      P   Rhat
sex:strata(proc)R           0.2554 0.1349 -0.0120  0.5244 0.0607 1.0008
sex:strata(proc)Tdth        0.1896 0.1728 -0.1474  0.5228 0.2727 1.0060
sex:strata(proc)Ttx         0.1463 0.1583 -0.1716  0.4564 0.3507 1.0084
ageD:strata(proc)R          0.0192 0.0310 -0.0412  0.0801 0.5313 1.0021
ageD:strata(proc)Tdth      -0.0011 0.0490 -0.0972  0.0941 0.9904 1.0114
ageD:strata(proc)Ttx        0.0638 0.0453 -0.0270  0.1501 0.1667 1.0040
value(lf):procR            -0.6926 0.2592 -1.1976 -0.1801 0.0089 1.0068
value(lf):procTdth         -0.7067 0.4394 -1.6262  0.1130 0.0929 1.0767
value(lf):procTtx          -1.4149 0.4522 -2.3014 -0.5480 0.0004 1.0856
vexpit(value(pa)):procR     0.2385 0.3841 -0.5141  0.9919 0.5144 1.0105
vexpit(value(pa)):procTdth  1.8997 0.8564  0.3158  3.6424 0.0213 1.0754
vexpit(value(pa)):procTtx   1.3061 0.8332 -0.3350  2.9684 0.1196 1.0803
frailty:strata(proc)Tdth   -0.0151 0.4568 -0.9993  0.8258 0.9969 1.0216
frailty:strata(proc)Ttx     0.5288 0.3994 -0.1668  1.4015 0.1400 1.0228
...
\end{Verbatim}

\noindent The \texttt{Survival Outcome} block of the \texttt{summary()} output now contains process-specific regression coefficients and association parameters on the log-hazard ratio scale.
\begin{sloppypar}
The frailty association coefficients quantify the residual dependence between the recurrent process and each terminal event beyond what is explained by the longitudinal markers and observed covariates. For example, \texttt{frailty:strata(proc)Tdth} corresponds to $\alpha_{F\mathrm{Tdth}}$ and represents the log-hazard ratio for death associated with a unit increase in the shared frailty $\upsilon_i$. Their relative magnitude reflects whether the shared frailty is more strongly associated with one terminal event or the other. When either of these coefficients is close to zero, the frailty adds little to the corresponding cause-specific hazard beyond what is already explained by the longitudinal process and covariates.
\end{sloppypar}
The current implementation in \textbf{JMbayes2} also allows recurrent events to be combined with a multistate process. In this setting, the event-time data should be organized in a single start--stop stacked dataset, with each row representing either a recurrent-event at-risk interval or a transition-specific risk interval, and a stratum variable distinguishing the recurrent-event process from the different multistate transitions.

\section{Conclusion}

Shared-parameter joint models provide a framework for analyzing measurements of the same individuals collected over time and intrinsically related event processes. They achieve this by linking mixed-effects submodels for the longitudinal outcomes with proportional hazards submodels for the event-time processes through shared latent random effects, allowing event hazards to depend on clinically meaningful functionals of the subject-specific longitudinal trajectories. This tutorial provided a practical, step-by-step guide to fitting and interpreting extended shared-parameter joint models for longitudinal and event-time data using the \textsf{R} package \textbf{JMbayes2}. Using simulated data resembling a real-world dataset, we progressively extended the framework to accommodate multiple longitudinal outcomes of different distributions, richer association structures, competing risks, multistate processes, recurrent events, %left truncation, %\hly{interval censoring}, 
and combinations thereof.

In applied analysis, most difficulties arise from practical rather than conceptual issues. Careful construction of the longitudinal and event-time datasets is essential; it is important to ensure that subject identifiers are consistent across datasets, that longitudinal and event-time variables share a common time scale and time origin, and that event indicators are coded correctly, especially when converting data into start--stop or stacked formats. For model fitting, practitioners should start with parsimonious specifications, check MCMC convergence, and then increase the complexity gradually. Association parameters should be interpreted on the scale implied by the link function and the chosen functional form. When multiple longitudinal outcomes are included, association parameters should be interpreted conditionally on the other outcomes and covariates, as mutually adjusted contributions.

Once convergence is established, it is important to assess whether the fitted model adequately describes the data before drawing substantive conclusions. For longitudinal processes, observed measurements can be compared with fitted subject-specific values. Longitudinal residuals can also be inspected for systematic patterns over time or across fitted values. In addition, posterior predictive checks offer a principled Bayesian approach to evaluating whether the fitted model adequately reproduces important features of the observed data across submodels and their associations~\citep{rizopoulos2026goodness}. These checks help identify which model components may be misspecified and motivate extensions such as nonlinear trajectories or alternative association structures.

Joint models, especially those involving multiple longitudinal outcomes, event processes, or large sample sizes, can be computationally demanding. When this becomes a limiting factor, strategies include simplifying covariance structures and exploiting scalable Bayesian estimation strategies such as the split-consensus approach proposed by \citet{miranda2026evaluating}. This method, which is available in \textbf{JMbayes2}, can substantially reduce wall-clock computation time while producing estimates that closely approximate those from a full-data analysis.

\textbf{JMbayes2} is under active development and its capabilities continue to expand. The package's website at \url{https://drizopoulos.github.io/JMbayes2} includes vignettes on topics not covered in this tutorial, such as baseline hazard specification, causal effects, time-varying effects, dynamic prediction, and the evaluation of predictive performance and clinical utility. These resources provide additional examples and practical advice, as well as updates on newly supported model features and support functions as \textbf{JMbayes2} continues to evolve. Readers seeking a comprehensive treatment of the statistical theory underlying joint models are referred to \citet{rizopoulos2012joint} and, for a Bayesian perspective, to \citet{rizopoulos2011bayesian}.

\bibliographystyle{plainnat}
\bibliography{references}

\newpage

\setcounter{table}{0}
\renewcommand{\thetable}{S\arabic{table}}

\setcounter{figure}{0}
\renewcommand{\thefigure}{S\arabic{figure}}

\setcounter{equation}{0}
\renewcommand{\theequation}{S\arabic{equation}}

\section*{Supplementary Material}

\subsection*{A. Prior Specification} \label{sup:prior}

Table~\ref{tab:mth:prior} summarizes the default priors for the models considered in this tutorial, as implemented in \textbf{JMbayes2}. The coefficient priors are specified on the scale used internally for estimation. The superscript $(s)$ denotes an estimate from the separately fitted submodel. Tildes denote the longitudinal fixed effects after centering, the survival regression and association coefficients after standardization of their design matrices, and the corresponding spline coefficients for the log baseline hazards. 

For the centered longitudinal fixed effects $\widetilde{\boldsymbol\beta}_j$, the default prior
covariance is
\begin{equation*}
\mathbf V_{\beta_j}
=\operatorname{diag}(v_{j1},\ldots,v_{jp_j}),
\qquad
v_{jr}=\min\!\left\{14400\,V_{j,rr}^{(s)},\,1000\right\},
\end{equation*}
where $p_j$ is the number of fixed effects and $\mathbf V_j^{(s)}$
is their estimated covariance matrix after centering.

For the longitudinal random effects $\mathbf b_i$,
we decompose their covariance matrix as $\mathbf D=\mathbf S_b\mathbf R\mathbf S_b$, where
$\mathbf S_b=\operatorname{diag}(s_{b1},\ldots,s_{bq})$ and
$q=\dim(\mathbf b_i)$.
The Lewandowski--Kurowicka--Joe (LKJ) prior is assigned to the
correlation matrix $\mathbf R$, with separate gamma priors for
the standard deviations $s_{br}$.

For the spline coefficients of the log baseline hazards,
the Gaussian difference-penalty prior is specified by the kernel
\begin{equation*}
p\!\left(\widetilde{\boldsymbol\gamma}_{0k}\mid\tau_k\right)
\propto
\tau_k^{r_k/2}
\exp\!\left\{
-\frac{\tau_k}{2}
\widetilde{\boldsymbol\gamma}_{0k}^{\top}
\mathbf M_k
\widetilde{\boldsymbol\gamma}_{0k}
\right\},
\end{equation*}
where
$\mathbf M_k=\boldsymbol\Delta_{k,u}^{\top}\boldsymbol\Delta_{k,u}$,
$\boldsymbol\Delta_{k,u}$ is the matrix of $u$th-order differences
($u=2$ by default), and $r_k=\operatorname{rank}(\mathbf M_k)$.

\begin{table}[h!]
\centering
\caption{Default prior specifications for the parameter blocks
used in the tutorial. For normal distributions, the second
argument denotes the variance or covariance matrix. Gamma
distributions use the shape--rate parameterization. Abbreviations: LKJ, Lewandowski--Kurowicka--Joe.}
\label{tab:mth:prior}

\begin{tabular}{lll}
\toprule
\textbf{Component} & \textbf{Parameter} & \textbf{Prior} \\
\midrule

Longitudinal submodels & & \\
& $\widetilde{\boldsymbol\beta}_j$
& $\mathcal N\!\left(
    \widetilde{\boldsymbol\beta}_j^{(s)},
    \mathbf V_{\beta_j}\right)$ \\

& $\sigma_{y_j}$
& $\Gamma\!\left(5,\,5/\sigma_{y_j}^{(s)}\right)$ \\

Shared random effects & & \\
& $\mathbf R$
& $\operatorname{LKJ}(3)$ \\

& $s_{br}$
& $\Gamma\!\left(5,\,5/s_{br}^{(s)}\right)$ \\

Event-time submodels & & \\
& $\widetilde{\boldsymbol\gamma}_{0k}\mid\tau_k$
& (see text) \\

& $\tau_k$
& $\Gamma(5,\,0.5)$ \\

& $\widetilde{\boldsymbol\gamma}_k$
& $\mathcal N(\boldsymbol\gamma^{(s)}_k,\,4\mathbf I)$ \\

& $\widetilde{\boldsymbol\alpha}$
& $\mathcal N(\mathbf0,\,4\mathbf I)$ \\

Frailty & & \\
& $\sigma_\upsilon$
& $\Gamma(0.625,\,2.5)$ \\

& $\alpha_{Fk}$
& $\mathcal N(0,\,4)$ \\

\bottomrule
\end{tabular}
\end{table}

\clearpage

\subsection*{B. Model Parameters}

\begin{table}[h!]
\centering
\caption{Main parameter blocks stored in the \texttt{mcmc}
component of fitted \texttt{jm()} objects.}
\label{tab:supp:model_params}
\renewcommand{\arraystretch}{1.15}
\begin{tabular}{p{3.2cm}p{8cm}}
\toprule
\textbf{Name} & \textbf{Description} \\
\midrule
\texttt{betas} or \texttt{betas1}, \texttt{betas2}, \ldots &
Fixed-effect coefficients for the longitudinal submodels, stored separately for each outcome. \\
\texttt{sigmas} &
Residual standard deviation(s) or other family-specific
dispersion or shape parameters (when applicable).\\
\texttt{D} &
Elements of the random-effects covariance matrix. \\
\texttt{gammas} &
 Regression coefficients for the baseline or exogenous covariates in the event-time submodel(s). \\
\texttt{alphas} &
Association coefficient(s) for the longitudinal functional(s) forms in the event-time submodel(s). \\
\texttt{bs\_gammas} &
Spline coefficients for the log baseline hazard(s) (when applicable). \\
\texttt{tau\_bs\_gammas} &
Smoothing precision parameters for the spline coefficients of the log baseline hazards (when applicable). \\
\texttt{frailty} &
Subject-specific frailty (when applicable). \\
\texttt{alphaF} &
Frailty association coefficient(s) (when applicable). \\
\texttt{sigmaF} &
Standard deviation of the frailty distribution (when applicable). \\
\texttt{b} &
Subject-specific random effects. Full posterior draws are retained only when
\texttt{save\_random\_effects = TRUE}. \\
\bottomrule
\end{tabular}
\end{table}

\end{document}